# Microwave dynamics of high aspect ratio superconducting nanowires studied using self-resonance

Daniel F. Santavicca,[1*] Jesse K. Adams,[1] Lierd E. Grant,[1] Adam N. McCaughan,[2] and Karl K. Berggren[2]

1. Department of Physics, University of North Florida, Jacksonville, FL 32224, USA
2. Department of Electrical Engineering and Computer Science, Massachusetts Institute of Technology, Cambridge, MA 02139, USA

*daniel.santavicca@unf.edu

## Abstract

We study the microwave impedance of extremely high aspect ratio (length/width ≈ 5,000) superconducting niobium nitride nanowires. The nanowires are fabricated in a compact meander geometry that is in series with the center conductor of a 50 Ω coplanar waveguide transmission line. The transmission coefficient of the sample is measured up to 20 GHz. At high frequency, a peak in the transmission coefficient is seen. Numerical simulations show that this is a half-wave resonance along the length of the nanowire, where the nanowire acts as a high impedance, slow wave transmission line. This resonance sets the upper frequency limit for these nanowires as inductive elements. Fitting simulations to the measured resonance enables a precise determination of the nanowire's complex sheet impedance at the resonance frequency. The real part is a measure of dissipation, while the imaginary part is dominated by kinetic inductance. We characterize the dependence of the sheet resistance and sheet inductance on both temperature and current and compare the results to recent theoretical predictions for disordered superconductors. These results can aid in the understanding of high frequency devices based on superconducting nanowires. They may also lead to the development of novel superconducting devices such as ultra-compact resonators and slow-wave structures.

**I. Introduction**

Kinetic inductance arises from the storage of kinetic energy in moving charge carriers. The kinetic inductance of a conductor is inversely proportional to the carrier density and the cross-sectional area. In normal conductors, this means that a large kinetic inductance is associated with a large resistance. In a superconductor, which has zero resistance at low frequency, this is not the case. A nanowire made of a low carrier density superconductor can have a kinetic inductance that is several orders of magnitude larger than its magnetic inductance. Such a nanowire enables the realization of relatively large inductances with very small dissipation in a compact planar geometry.[1] Kinetic inductance depends nonlinearly on both current and temperature, resulting in a number of promising device applications. It is important for understanding the performance of microwave kinetic inductance detectors (MKIDs),[2] superconducting nanowire single-photon detectors (SNSPDs),[3] superconducting parametric amplifiers,[4] superconducting metamaterials,[5] and superconducting resonators.[6]

High inductance superconducting nanowires have also been explored as a component in superconducting qubits.[7,8] Related to this application is the concept of superinductance,[9,10] which is a nondissipative microwave impedance whose magnitude is larger than the resistance quantum $h/(2e)^2$ = 6.45 kΩ. Such a superinductance is effective at suppressing low-frequency charge fluctuations, a key source of decoherence in superconducting qubits. Theoretically, the impedance of an inductor becomes infinite at infinite frequency. In practice, there is always some maximum frequency beyond which parasitic capacitance will cause the impedance to deviate significantly from ideal inductive behavior. This effect determines the upper frequency limit, and hence the upper impedance limit, for an inductor.

Measurements of the high frequency impedance of superconducting nanowires allow us to determine the frequency beyond which they no longer look like simple inductive elements. To facilitate accurate high frequency measurements, we fabricated high aspect ratio superconducting nanowires in a compact meander geometry. This meander is in series with the center conductor of a 50 Ω coplanar waveguide (CPW) transmission line. The nanowire meander is the same geometry typically used in an SNSPD, although the placement in the CPW is distinct from typical SNSPD device architecture. We measure the transmission coefficient (the ratio of transmitted voltage to incident voltage) of the CPW as a function of frequency up to 20 GHz.

At low frequencies, the nanowire behaves approximately as an ideal inductor. At higher frequencies, we see a first-order resonance whose signature is a peak in the magnitude of the transmission coefficient. This resonance sets the upper frequency limit for these nanowires as inductive elements. The frequency of the resonance depends on the temperature and DC bias current.

We compare the measured resonance to numerical simulations of the device geometry in AWR Microwave Office. The capacitance is determined by the geometry, and the sheet inductance of the nanowire is used as a fitting parameter to match the simulated resonant frequency to the measured resonant frequency. As the inductance of our samples is dominated by kinetic inductance, this procedure allows us to determine the temperature- and current-dependence of the kinetic inductance. We compare these results to recent calculations for disordered superconductors by Clem and Kogan.[11]

The quality factor of the resonance is highly sensitive to dissipation. In the simulations, we adjust the real part of the nanowire sheet impedance (the sheet resistance) to match the measured resonance peak height. This allows us to determine the dependence of the resistance on temperature and current. Again, comparison is made to the theory developed by Clem and Kogan.[11]

Simulations of the nanowire current distribution at the first-order resonance show that this is a half-wave standing wave resonance. Because of its extremely large inductance per unit length, the nanowire acts as a high impedance, slow wave transmission line in which electrical signals propagate ≈25 times slower than in free space.[12] The extreme compression of the wavelength on the nanowire compared to the free space wavelength means that the nanowire must be treated as a distributed element rather than a lumped element at frequencies that are lower than might be expected for a sample of this size. Such a distributed element description has important implications for understanding the dynamics of devices incorporating high aspect ratio nanowires. It also presents interesting possibilities for the miniaturization of superconducting resonators and for the development of slow-wave structures.

**II. Experiment**

To fabricate the sample chip, first niobium nitride (NbN) was deposited on a 500 µm thick R-plane sapphire substrate using a DC magnetron sputtering system, which sputtered a niobium target with a power of 91 W in a 26:8 Ar:$N_2$ atmosphere at a pressure of 0.33 Pa (2.5 mTorr) for 100 s. These parameters produced a NbN film approximately 5 nm thick with a sheet resistance of 498 Ω/sq. Once the NbN film was deposited, gold contact pads were added to the surface using a photolithographically-defined liftoff process. The liftoff process began by spinning Shipley S1813 photoresist on the wafer at 5 krpm for 60 s, baking the wafer at $90^0$ C for 2 min, and then exposing the contact pads pattern using a mask and UV flood exposure tool. After exposing the photoresist to 60 mJ/$cm^2$ of UV light through the mask, the sample was developed in CD-26 for 12 s, and the resulting patterned sample was placed in an electron-beam evaporation tool which deposited a 10 nm titanium adhesion layer followed by 100 nm of gold. The sample was then sonicated in an acetone bath for 3 min to complete the liftoff and remove the unwanted photoresist and gold.

Once the sample had gold contact pads, nanowires were fabricated from the NbN film using an electron-beam lithography process. To begin, 4% HSQ was spun on the NbN surface at 5 krpm for 60 s. The nanowires were then written using an Elionix electron-beam lithography tool at 125 kV and a beam current of 900 pA with an areal exposure dosage of 3460 µC/$cm^2$. After exposing the sample, the HSQ was developed in 25% TMAH for 3 min, and the resulting nanowire patterns were etched into the NbN using a $CF_4$ RIE process. The result is a nanowire with a width of 100 nm that is in a meander geometry consisting of 33 straight segments of 14 μm length. Adjacent segments are separated by 200 nm, and the connecting sections at each end are curved to minimize current crowding.[13] The total nanowire length is approximately 0.5 mm. The nanowire geometry can be seen in Fig. 1.

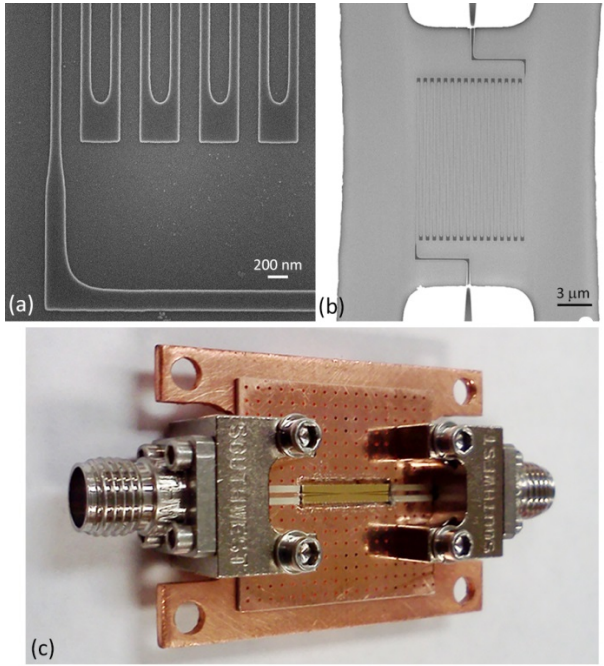

Fig. 1. (a) Scanning electron micrograph of a section of the NbN nanowire. (b) Scanning electron micrograph showing nanowire and center conductor of the CPW structure. (c) Photograph of sample chip mounted in microwave sample holder (sample holder cover not shown).

Each chip contains five devices, four with nanowires and one that is a continuous CPW (no nanowire) which serves as a normalization sample. Prior to measuring, the chip is diced in order to separate each of the five devices into separate chips of approximately 1.8 mm × 10 mm. Separating the individual devices is important for efficient high-frequency coupling, as it enables a robust connection between the chip ground and the sample holder ground. A device is mounted in a sample holder, as seen in Fig. 1(c). The sample holder has a 50 Ω ground-backed CPW transmission line manufactured from Rogers RT/duroid 6010.2 LM low-loss dielectric. This material has a dielectric constant of 10.2, chosen to approximately match the dielectric constant of the sapphire device substrate. An array of copper-plated vias connect the top and bottom sample holder ground planes. The sample is connected via 25 μm diameter aluminum wirebonds. The bond length is kept short, < 0.5 mm, and five to six bonds are used on each side of the center conductor of the CPW to minimize excess inductance. Approximately 100 wirebonds at regular intervals connect the chip ground to the sample holder ground. The ground-backed CPW is mounted on a solid copper base and an SMA end-launch connector (Southwest Microwave 292-04A-5) is connected to each end. A copper lid covers the sample holder.

The sample holder is mounted in a liquid helium cryostat which has a base temperature of 1.5 K. SMA-connectorized semirigid coaxial lines with a stainless steel outer conductor and a copper inner conductor connect the sample to room temperature. A bias-tee separates the RF and DC signals, and an ≈ 10 Hz low-pass filter is used on the DC port. Typical RF input power incident on the nanowire is -90 dBm; care was taken to ensure that the probe power was sufficiently small so as not to affect the measured resonances. (At higher microwave power, the nonlinear current-dependence of kinetic inductance and microwave-induced dissipation can change the shape of the resonance.) The output signal is amplified with three stages of amplification at room temperature. The first stage is an 8-18 GHz amplifier with a gain of 21 dB and a noise figure of 2.2 dB. The next two stages provided additional gain of approximately 56 dB.

To calibrate the experiment and to ensure that there are no unwanted sample holder resonances, we mounted one of the normalization samples (a continuous CPW with no nanowire) and measured its transmission coefficient $S_{21} = 20\log(V_{out}/V_{in})$, where $V_{out}$ is the output voltage and $V_{in}$ is the input voltage, at the base temperature of our cryostat (1.5 K). This normalization measurement is shown in Fig. 2. Also shown are the raw data taken with a nanowire sample showing a resonance at 12.5 GHz. Each nanowire transmission measurement is then normalized by dividing by the measured transmission coefficient of the continuous CPW sample. This normalization technique provides an effective way to eliminate contributions to the measured response from everything other than the nanowire.

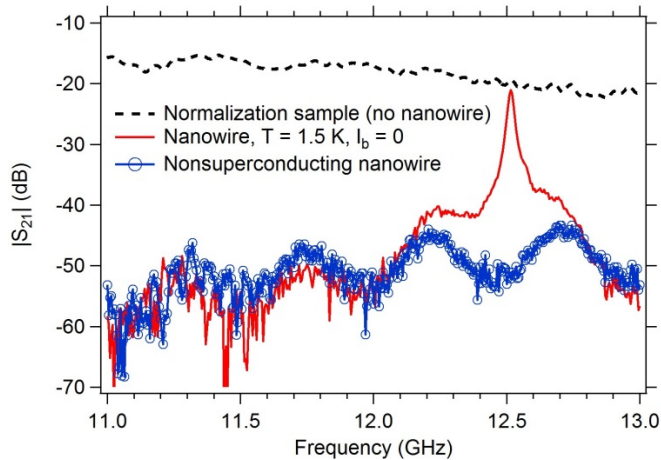

**Fig. 2.** Measured magnitude of the transmission coefficient $|S_{21}|$ with nanowire sample at 1.5 K and zero bias current. Also shown are the measured transmission coefficient with a continuous CPW (no nanowire), which is used to normalize the nanowire data, and the measured transmission coefficient with the nanowire in the nonsuperconducting state ($T$ = 1.5 K and a bias current that exceeds the switching current).

Two sample chips were fabricated containing a total of eight nanowire devices and two normalization samples. After inspection in an optical microscope and probing the room temperature DC resistance, five nanowire devices were selected for low temperature measurement. All five displayed similar critical temperatures (≈ 11 K) and clear resonances at similar frequencies (≈ 12 GHz at low temperature). All samples showed a similar decrease in the resonance frequency as the temperature was increased. The biggest variation between the devices was in the measured switching current as well as the decrease in the resonant frequency as the DC bias current was increased. This is consistent with the switching current being reduced from the depairing current due to constrictions, a phenomenon commonly observed in this type of NbN nanowire.[14] In this paper, all data presented are from the nanowire sample that showed both the highest switching current as well as the largest increase in kinetic inductance with increasing bias current, consistent with having the lowest level of constriction. DC resistance versus temperature and current versus voltage data for this sample are shown in Fig. 3.

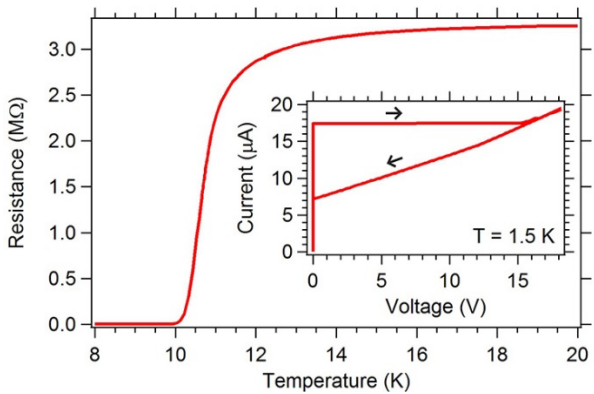

Fig. 3. Resistance versus temperature measured with 10 nA DC current. Inset: Current-voltage curve measured at 1.5 K. Directions of the voltage sweep are indicated with arrows.

### III. Results

The measured transmission coefficient of the nanowire sample, normalized as described previously, is compared to simulations of the device geometry in AWR Microwave Office. The simulations describe the nanowire with a complex impedance defined in terms of a sheet resistance and sheet reactance. We assume that the sheet reactance is dominated by kinetic inductance. The contribution to the sample reactance from the geometric capacitance is captured through the simulation geometry.

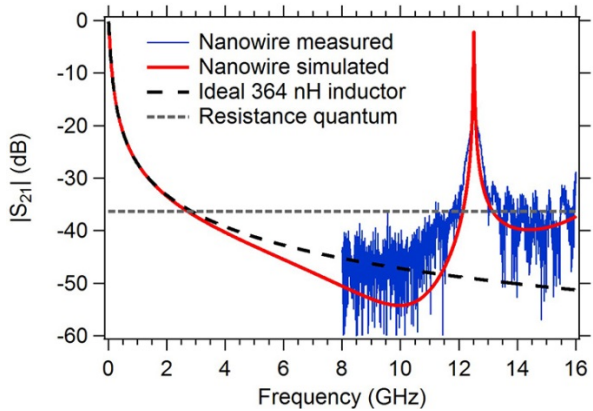

Fig. 4. Measured magnitude of the transmission coefficient $|S_{21}|$ for nanowire sample at $T$ = 1.5 K and zero bias current. Simulation is for a sheet inductance of 78.15 pH/sq and a sheet resistance of 2 mΩ/sq, chosen to match the data. Also shown for comparison is the result for an ideal 364 nH inductor and the resistance quantum $h/(2e)^2$. We see that the nanowire impedance begins to exceed the resistance quantum above 2.7 GHz; this is the superinductance regime.

In Fig. 4 we plot a simulation of the magnitude of the transmission coefficient $|S_{21}|$ assuming a sheet resistance of 2 mΩ/square and a kinetic inductance of 78.15 pH/square, which was determined by matching the simulated resonant frequency to a measurement performed at zero bias current and a temperature of 1.5 K. (The magnetic inductance is ~ 1 nH and hence can be safely neglected.) Shown for comparison is $|S_{21}|$ for an ideal 364 nH inductor. The nanowire simulation closely follows that of the ideal inductor up to approximately 4 GHz. Also shown for reference is $|S_{21}|$ for an ideal resistor equal to the resistance quantum $R_Q$ = $h/(2e)^2$ = 6.45 kΩ. (We use $2e$ as the Cooper pair charge.) The frequency range over which the nanowire $|S_{21}|$ is lower than that of $R_Q$ corresponds to the superinductance regime.[9,10]

In order to understand the nature of the first-order resonance, we simulated the current distribution on the nanowire at the first-order resonant frequency. The result is shown in Fig. 5. We see that this resonance corresponds to a half-wavelength standing wave along the length of the nanowire, with a current minimum in the center and current maxima at each end of the nanowire. This is similar to a simulation presented in Ref. 1.

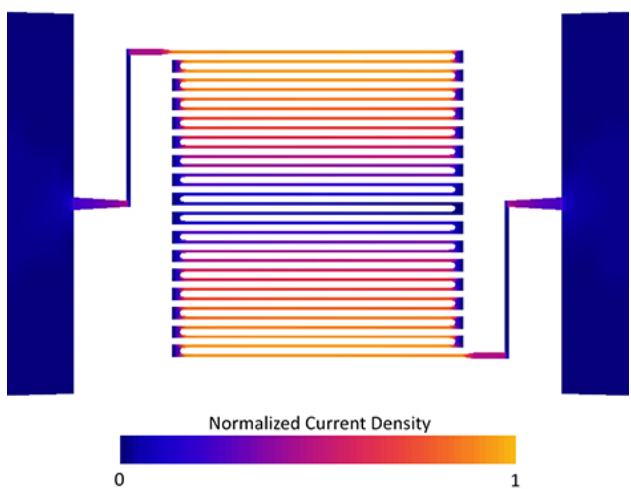

Fig. 5. Simulation of the normalized current density at the first-order resonance (12.5 GHz). (The CPW ground is not seen here but is present in the simulation.)

Given that this is a half-wave resonance, we can easily determine the wavelength on the nanowire as well as the propagation velocity at the resonance frequency. From the data in Fig. 4 (zero DC bias current and a temperature of 1.5 K), we find a wavelength at the 12.5 GHz resonance of $\lambda$ = 1.0 mm. This is 4.2% of the free space wavelength. The propagation velocity, $v_P = \lambda f = 1.25 \times 10^7$ m/s, is also 4.2% of the free space value. We see that the nanowire behaves as a high impedance, slow wave transmission line. If we extend the simulation to higher frequencies (not shown), we see higher order resonances that are not simply at integer multiples of the first-order resonance frequency. This demonstrates that the higher-order resonances are more complex, which may arise from capacitive coupling of the signal through adjacent lines in the meander geometry. We focus our present study on the first-order resonance.

Increasing either the temperature or the DC bias current results in the resonance shifting to lower frequency, consistent with an increase in kinetic inductance. This behavior is shown in Fig. 6. The width of the resonance also increases (the quality factor decreases) and the peak height decreases due to increasing dissipation. The simulated resonance is matched to each measured resonance by adjusting the sheet resistance and sheet reactance in the simulation. We were able to achieve very good agreement between the simulation and the measured data close to the resonant frequency. Farther from the resonant frequency, the agreement was not as good, which may be due to the decreased signal-to-noise ratio. This procedure allows us to study the dependence of both the resistance and kinetic inductance on temperature and current.

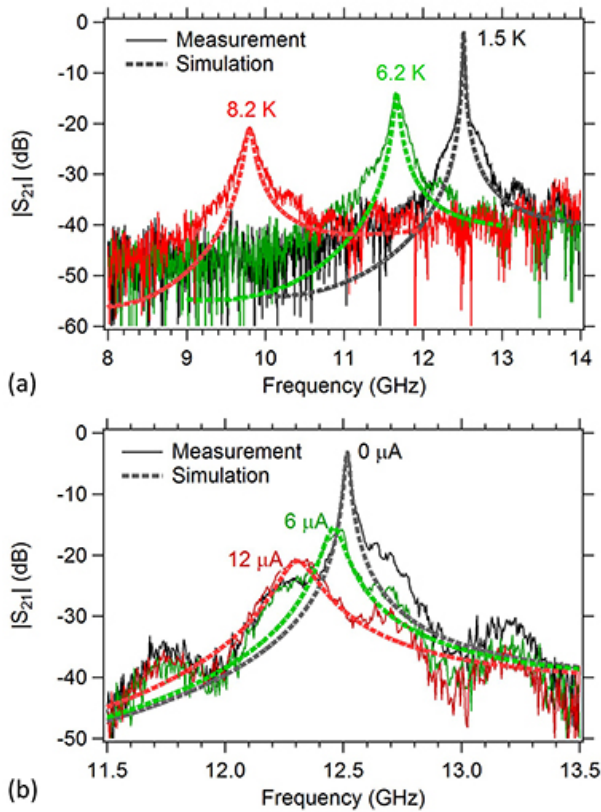

Fig. 6. Comparison of measured and simulated resonances at (a) three different temperatures with zero bias current, and (b) three different bias currents at a temperature of 1.5 K.

## A. Temperature dependence

The kinetic inductance $L_K$ of a wire of length $l$, width $w$, and thickness $d$ can be expressed in terms of the superconducting carrier density $n_s$ as[1]

$$L_K = \left(\frac{m}{2n_s e^2}\right)\left(\frac{l}{wd}\right) \quad (1)$$

where $m$ is the free electron mass and $e$ is the fundamental charge. The complex conductivity of the nanowire is $\sigma = \sigma_1 - i\sigma_2$. In the low frequency limit ($hf \ll kT$), the Mattis-Bardeen expression for the ratio of $\sigma_2$ to the normal state conductivity $\sigma_N$ is given by[15]

$$\frac{\sigma_2}{\sigma_N} = \frac{\pi\Delta(T)}{hf}\tanh\left(\frac{\Delta(T)}{2kT}\right) \quad (2)$$

where $\Delta(T)$ is the temperature-dependent superconducting energy gap. $\Delta(0) = 1.76kT_C$, and the temperature-dependence of the energy gap $\Delta(T)$ can be solved numerically as described in Refs. 11 and 15. Assuming that the imaginary component of the conductivity is dominated by kinetic inductance, we can write the ratio of the finite-temperature kinetic inductance $L_K(T)$ to the zero-temperature kinetic inductance $L_K(0)$ as

$$\frac{L_K(T)}{L_K(0)} = \frac{\Delta(0)}{\Delta(T)\tanh\left(\frac{\Delta(T)}{2kT}\right)}. \quad (3)$$

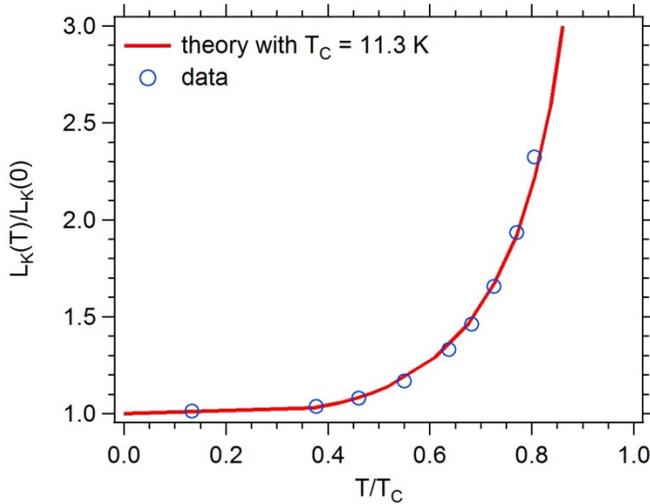

Fig. 7. Normalized kinetic inductance versus normalized temperature. The theory is from eq. 2 with $T_C$ as an adjustable parameter. The best agreement is achieved with $T_C$ = 11.3 K. The experimental $T_C$ is 10.7 K. The zero temperature sheet inductance is 78 pH per square.

In Fig. 7 we plot the experimental values for the kinetic inductance normalized by its value at the lowest temperature measured (1.5 K), which is taken as equivalent to the zero temperature result. These values are determined from the fitting procedure shown in Fig. 6. The uncertainties in these values are in most cases less than the size of the data points in Fig. 7. We also plot the predicted behavior from Eq. 3. We use $T_C$ as an adjustable parameter and find good agreement between the theory and the data using a $T_C$ of 11.3 K in the theory. This is slightly higher than the measured critical temperature of 10.7 K, defined as the temperature at which the nanowire has half of its normal state resistance at 20 K. We are not able to measure above $t = T/T_c > 0.82$ because the resonance quality factor becomes too low to accurately assign a resonance frequency.

The sheet resistance $R_S$ is related to the complex conductivity by

$$R_S = \frac{1}{d}\text{Re}\left[\frac{1}{\sigma}\right] = \frac{1}{d}\left(\frac{\sigma_1}{\sigma_1^2 + \sigma_2^2}\right) \approx \frac{1}{d}\left(\frac{\sigma_1}{\sigma_2^2}\right) \tag{4}$$

where $d$ is the film thickness and the approximation assumes $\sigma_2 \gg \sigma_1$, which is the case for high inductance superconducting nanowires such as the ones considered here.

In BCS theory, $\sigma_1$ is found from[11,15]

$$\frac{\sigma_1}{\sigma_N} = \frac{1}{\omega}\int_{-\infty}^{\infty} d\omega' \left[F\left(\hbar\omega' - \frac{\hbar\omega}{2}\right) - F\left(\hbar\omega' + \frac{\hbar\omega}{2}\right)\right] \times \\ \left[n\left(\hbar\omega' - \frac{\hbar\omega}{2}\right)n\left(\hbar\omega' + \frac{\hbar\omega}{2}\right) + p\left(\hbar\omega' - \frac{\hbar\omega}{2}\right)p\left(\hbar\omega' + \frac{\hbar\omega}{2}\right)\right] \tag{5}$$

where $\omega$ is the angular measurement frequency, $F = \left(1 + e^{\hbar\omega/kT}\right)^{-1}$ is the Fermi function and, for the pair-breaking regime applicable here, $n = \text{Re}\left[u/\sqrt{u^2 - 1}\right]$ and $p = \text{Re}\left[1/\sqrt{u^2 - 1}\right]$ with $u$ defined by

$$\frac{\hbar\omega}{\Delta} = u\left(1 - i\frac{\zeta}{\sqrt{u^2 - 1}}\right) \tag{6}$$

and $\zeta = D\kappa^2/(2\Delta)$ with $D$ the diffusion coefficient and $\kappa$ the phase gradient. See Ref. 11 for details on the evaluation of Eq. 5. At zero bias current, $\zeta = 0$. We use numerical integration to evaluate Eq. 5. This ratio is then used with Eqs. 2 and 4 to calculate the sheet resistance as a function of temperature. A comparison of this calculation with the sheet resistance determined from fitting to the measured resonance peaks is shown in Fig. 8. The normal state resistance is used as an adjustable parameter in the calculation; the best agreement with the data is obtained for a normal state resistance of 1.0 MΩ, while the DC normal state resistance (measured at 20 K) is approximately 3 MΩ. This discrepancy could be due in part to uncertainty in the film thickness, although a full explanation for the discrepancy is presently lacking.

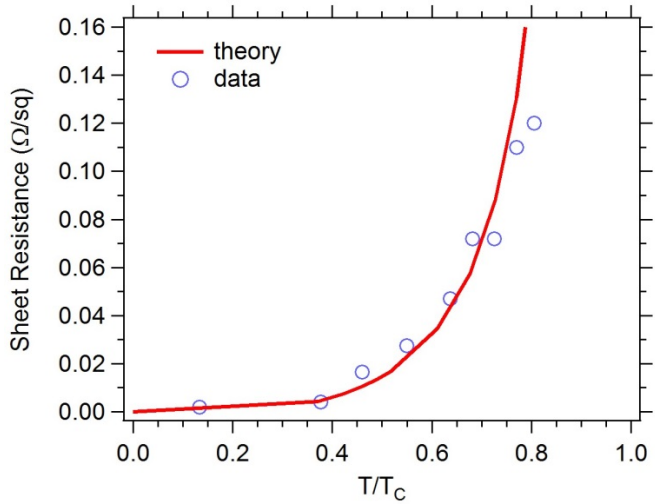

Fig. 8. Sheet resistance versus normalized temperature.

To emphasize the picture of the nanowire as a high impedance, slow wave transmission line, we plot in Fig. 9 the wavelength on the nanowire $\lambda$ normalized by the free space wavelength as a function of normalized temperature. The wavelength is found from the measured first-order resonance, using the fact that this is a half-wave resonance. This ratio is the same as the ratio of the propagation velocity of the electromagnetic wave on the nanowire $v_P = f\lambda$ normalized by the free space velocity. Also plotted is the characteristic impedance of the nanowire $Z_C = \sqrt{L/C}$ versus normalized temperature. The nanowire's capacitance per unit length, which can be calculated from the propagation velocity and the inductance, is 8.2 pF/m.

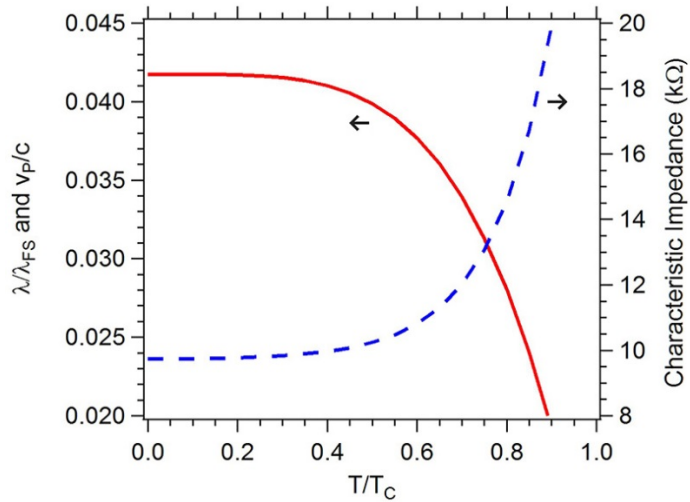

Fig. 9. Ratio of the wavelength on the nanowire $\lambda$ to the free space wavelength $\lambda_{FS}$, which is equal to the ratio of the propagation velocity on the nanowire $v_P$ to the speed of light in free space $c$, (left axis) and characteristic impedance of the nanowire (right axis) versus normalized temperature (using $T_C$ = 11.3 K).

## B. Current dependence

The current-dependence of kinetic inductance in the small current limit ($I \ll I_C$) is often approximated as quadratic.[4] In Ref. 11, the full current-dependence of the kinetic inductance at finite temperature is calculated in two different regimes: the slow relaxation (fast measurement) regime, in which the inverse of the measurement frequency is small compared to the relaxation time of the superconducting order parameter; and the fast relaxation (slow measurement) regime, in which the inverse of the measurement frequency is large compared to the relaxation time of the superconducting order parameter. The relaxation time of the order parameter is the timescale over which the superfluid density can change. For the slow relaxation case at $t = 0.2$, the calculated ratio of the current-dependent kinetic inductance $L_K(I)$ to the zero-current kinetic inductance $L_K(0)$ is described by the equation

$$L_K(I)/L_K(0) = 1.422 - (1.422-1)\left(1-(I/I_C)^{2.45}\right)^{1/2.45}$$

to an accuracy of 0.5%. For the fast relaxation case at $t = 0.2$, the calculated ratio is described by the equation

$$L_K(I)/L_K(0) = \left(1-(I/I_C)^{2.27}\right)^{-1/2.27}$$

to an accuracy of 1%. We note that $L_K(I)/L_K(0)$ depends only very weakly on temperature,[11] so the expression for $L_K(I)/L_K(0)$ at $t = 0.2$ should still be reasonably accurate for the data at $T = 1.5$ K ($t = 0.14$). In Fig. 10 we compare these two functions to the data for current-dependent kinetic inductance. The critical current is used as an adjustable parameter in each function. For the slow relaxation case (which predicts $L_K(I)/L_K(0) = 1.42$ as $I \to I_C$), we find the best agreement using a critical current of 25 µA. For the fast relaxation case (which predicts $L_K(I) \to \infty$ as $I \to I_C$), we find the best agreement using a critical current of 42 μA.

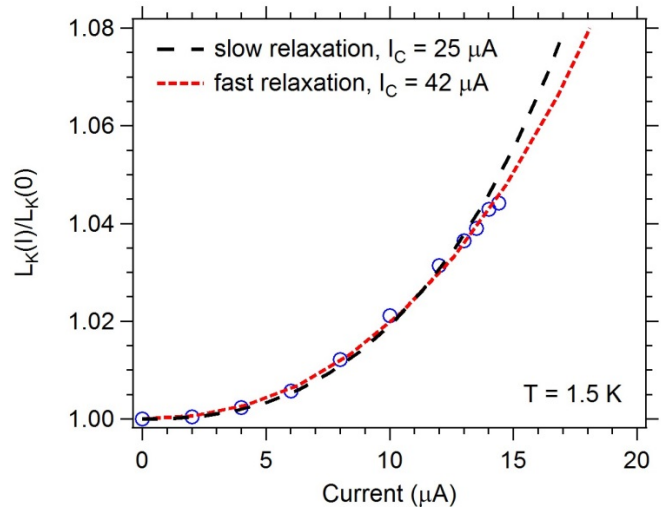

Fig. 10. Normalized kinetic inductance versus bias current at a temperature $T = 1.5$ K. Fits are performed for the slow and fast relaxation cases with the critical current $I_C$ used as an adjustable parameter. The best fits are achieved with $I_C = 25$ µA for the slow relaxation case and with $I_C = 42$ µA for the fast relaxation case. The measured critical current is 18 µA.

The measured switching current of this sample was 18 μA. It is likely that the measured switching current is reduced from the true depairing current due to constrictions.[14] Previous studies of similar NbN nanowires found typical measured critical current densities of 2–5 × 10$^{10}$ A/m$^2$ at low temperature ($T \ll T_C$),[3] with the higher end of this range corresponding to the least constricted samples. For our geometry, this range corresponds to critical currents of 10–25 μA, suggesting that the slow relaxation prediction is the appropriate one for this work. The slow relaxation case requires that the relaxation time of the superconducting order parameter be slow compared to 1/(12.5 GHz) = 80 ps. For very thin films at low temperature (1.5 K), we expect the relaxation time to be limited by electron-phonon scattering. Previous determinations of the electron-phonon scattering time in thin-film NbN found a value of ≈ 250 ps at 1.5 K with a temperature dependence of $T^{-1.6}$.[16]

Eq. 5 is used to calculate the current-dependence of $\sigma_1$, where in this case $\zeta$ is non-zero, following the procedure described in Ref. 11. We take $t$ = 0.2 and $hf = 0.02(1.76kT_c)$, where $f$ is the measurement frequency. $\sigma_1$ is then used with Eqs. 2 and 4 to calculate the sheet resistance. In Eq. 2, we use the current-dependent energy gap $\Delta(I)$ at $t$ = 0.2, calculated as described in Ref. 11.

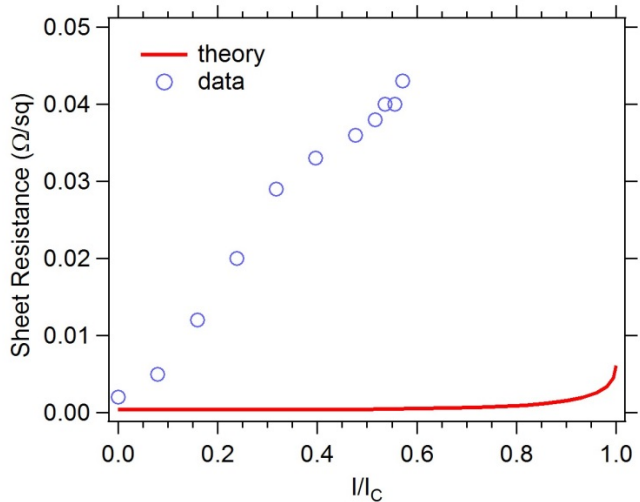

**Fig. 11. Experimental and calculated values for the sheet resistance as a function of the normalized current. The data were taken at $T$ = 1.5 K and the calculation assumes $t$ = 0.2. We assume a critical current of 25 μA.**

The result of this calculation is plotted along with the experimental data for the current-dependent sheet resistance in Fig. 11. There are no adjustable parameters in the calculation. We see significant disagreement between the experimental and the theoretical values.

At finite current, the measured sheet resistance is several orders of magnitude larger than the theory would suggest. Typical dark count rates in similar nanowires at 1.5 K are ~$10^2 – 10^4$ Hz,[17] with each dark count corresponding to the creation of a resistive region with a resistance of ≈ 6 kΩ for a duration of ≈0.25 ns.[18] The resulting time-average resistance due to dark counts is too low to explain the observed behavior. Another possibility is that the dissipation arises from trapped magnetic flux vortices in the nanowire, which could become mobile due to the Lorentz force from the bias current. The movement of the flux vortices would be a significant source of current-dependent dissipation.[19] This hypothesis will be explored in future studies.

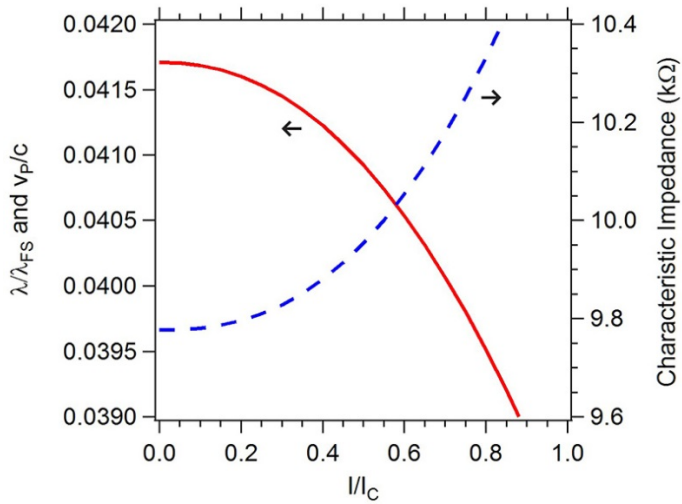

Fig. 12. Ratio of the wavelength on the nanowire $\lambda$ to the free space wavelength $\lambda_{FS}$, which is equal to the ratio of the propagation velocity on the nanowire $v_P$ to the speed of light in free space $c$, and characteristic impedance of the nanowire (right axis) versus normalized current (using a critical current of 25 μA).

Finally, we show in Fig. 12 the wavelength normalized by the free space value and the propagation velocity normalized by the free space value as a function of normalized current. Also plotted is the characteristic impedance of the nanowire as a function of normalized current. As in Fig. 9, these are found from the measured resonant frequency and the corresponding inductance determined from the simulations.

## IV. Conclusions

An accurate understanding of the microwave-frequency impedance of superconducting nanowires has important implications for a variety of devices such as SNSPDs, MKIDs, superconducting metamaterials, superconducting resonators, and superconducting qubits. We have characterized the microwave impedance of high aspect ratio superconducting NbN nanowires in a compact meander geometry. Simulations show that the first-order self-resonance is a half-wave resonance, where the nanowire behaves as a slow wave, high impedance transmission line. Matching numerical simulations to the measured resonance data allows us to determine the complex sheet impedance of the nanowire at the resonance frequency. The temperature- and current-dependence of the sheet resistance and sheet inductance are compared to predictions for disordered superconductors developed by Clem and Kogan.[11] Good agreement is found between the theory and the data for the temperature- and current-dependence of the kinetic inductance, where we assume that we are in the slow relaxation case for the current-dependence. Matching the theory and the data for the temperature-dependence of the sheet resistance requires assuming a factor of three smaller normal state resistance in the theory than is seen in DC measurements. A much larger discrepancy is seen between the theory and the data for the current-dependence of the sheet resistance. This may be due to the presence of flux vortices in the nanowire, although further study is needed to test this hypothesis. These results highlight the extreme compression of the electromagnetic wavelength and the propagation velocity on the nanowire that arises from its very large kinetic inductance. We believe these results may be significant in the development of novel superconducting devices based on high kinetic inductance materials.

**Acknowledgements**

We thank Y. Ban, J. Clem, J. Garner, and V. Kogan for helpful discussions. This work was supported by NSF grants ECCS-1509253 (UNF) and ECCS-1509486 (MIT). D.S. also acknowledges support from a Cottrell College Science Award from Research Corporation for Science Advancement.